\documentclass{iopart}

\usepackage{graphicx}

\usepackage{color}

\begin{document}

\title{Breaking the symmetry of a Brownian motor with symmetric potentials}

\author{H Hagman, M Zelan, C M Dion}
\address{Department of Physics, Ume{\aa} University, SE-901\,87
  Ume{\aa}, Sweden.} 
\ead{claude.dion@tp.umu.se}


\begin{abstract}
  The directed transport of Brownian particles requires a system with
  an asymmetry and with non-equilibrium noise. We here investigate
  numerically alternative ways of fulfilling these requirements for a
  two-state Brownian motor, realised with Brownian particles
  alternating between two phase-shifted, symmetric potentials. We show
  that, besides the previously known spatio-temporal asymmetry based
  on unequal transfer rates between the potentials, inequalities in
  the potential depths, the frictions, or the equilibrium temperatures
  of the two potentials also generate the required asymmetry. We also
  show that the effects of the thermal noise and the noise of the
  transfer's randomness depend on the way the asymmetry is induced.
\end{abstract}


\submitto{\JPA}


\section{Introduction}

Transport phenomena are ubiquitous in nature and are an interesting
physical problem. As the thermal noise becomes increasingly important
for the dynamics at shorter length scales, the control and the
theoretical treatment of the transport generally becomes more
complicated. Brownian motors do, however, take advantage of these
random fluctuations, as they channel them into useful energy in the
absence of bias
forces~\cite{RevModPhys.81.387,R.DeanAstumian05091997,BM1,BM2,BM3}. This
makes them interesting from the point of view of statistical physics,
and they have been proposed as the basis of a large number of
transport phenomena in biological
structures~\cite{RevModPhys.81.387,R.DeanAstumian05091997,Spec_appl_phys_A,%
  RevModPhys.69.1269,Mogilner:2003p1772}.

Brownian particles~\cite{Einstein_brownian} in a periodic potential
can be subjected to directed transport provided that two requirements
are fulfilled: (i) the system has to possess an asymmetry, in
accordance with the Curie principle~\cite{Curie}, \emph{i.e.}, the
trapping potentials have to present at least a spatial or
spatio-temporal asymmetry; (ii) the system has to be out of thermal
equilibrium, in agreement with the second law of
thermodynamics~\cite{rat_praw}.

In addition to the potentials, a friction force is usually present,
the combination of the two localising the particles close to the
potential minima, along with a diffusive force that is the source of
the Brownian motion.  This diffusive force is often treated by letting
the particles interact with a heat bath of temperature $T$. The
required asymmetry is usually included in the potential, \emph{e.g.},
as is the case for sawtooth potentials.  To satisfy the second
requirement stated above, the equilibrium between the particles and
the heat bath has to be broken for drifts to be induced. This is
usually done by non-adiabatically changing a parameter of the
potential, \emph{e.g.}, the spatial phase in a rocked ratchet or the
potential depth in a flashing ratchet, in a time-periodic
fashion~\cite{RevModPhys.81.387}.  The latter is equivalent to the
case where the particles can be found in different states, each with
its own characteristic static potential, and where it is the state
that evolves dynamically.

As an alternative to an asymmetric potential, it is possible to
achieved a biased drift for a multi-state particle using symmetrical
potentials that are properly shifted between the states (see,
\emph{e.g.},~\cite{Chen:1997p1597,brown:kanada99,Lee:2005p1611,Lee:2005p1612,RenzoniBM}). Directed
transport is then achieved through a combination of the shift of the
potentials and inequalities in the rates of change of the states,
resulting in a combined spatio-temporal asymmetry.  To understand the
processes involved, it can be convenient to consider this change of
state as a transfer from one potential to another.
 
Such a Brownian motor, where the asymmetry is given by a combination
of a non-zero relative spatial phase and unequal transfer rates
between only two symmetric, periodic potentials, has been demonstrated
both numerically~\cite{bm:sanchez-palencia04} and
experimentally~\cite{OurBM1,newBM,0295-5075-81-3-33001}, including
with real-time control of the drift~\cite{rtbm}.  The original idea of
using two phase-shifted symmetric
potentials~\cite{bm:sanchez-palencia04} (along with its
implementation~\cite{OurBM1,newBM,0295-5075-81-3-33001,rtbm}) was
based on laser-cooled atoms interacting with two optical lattices,
created from the interference of laser beams~\cite{Grynberg2001335}.
As such, it was assumed that the asymmetry coming from unequal
transfer rates between the two phase-shifted optical lattices, as
occurs ``naturally'' in the system, was the asymmetry driving the
Brownian motor~\cite{bm:sanchez-palencia04}.  However, this is far
from the only asymmetry present in the system.  As the frequency and
irradiance of the lasers making each optical lattice can be adjusted
independently, inequalities between the potentials depths, the
diffusion, and the friction are usually present and can be modified by
varying these parameters.

From a more general point of view, this two-state Brownian motor is a
complex system, as the symmetry can be broken by any inequality
between the two potentials, in combination with a relative spatial
phase shift. That is, instead of unequal transfer rates, inequalities
in the equilibrium temperature, the potential depth, or the friction
could be used. These multiple ways of breaking the symmetry have not
yet been investigated. Moreover, in the original system, the transfers
(or changes of state) are random, as they originate from spontaneous
emission.  This adds a source of randomness to the system, in addition
to the Brownian motion of the particles. We thus investigate below the
role of these two sources of fluctuations and of the different kinds
of asymmetries using numerical simulations.

\section{Model}%
The system considered consists of a Brownian particle interacting with
either of two symmetric potentials, 
\begin{eqnarray}
U_\mathrm{1} & = & \frac{A_1}{2}\left[1-\cos{(kx)}\right] ,
\nonumber\\ 
U_\mathrm{2} & = & \frac{A_2}{2}\left[1-\cos{(kx + \varphi)}\right] ,
\end{eqnarray}
where $A_i$ are the potential depths, $k$ is the wave number and
$\varphi$ the relative spatial phase of the potentials. We hereafter
scale length such that $k = 2\pi$.  The particle will be transferred
between the states (the potentials) with transfer rates
$\Gamma_\mathrm{1\rightarrow2} = 1/\tau_1$ and
$\Gamma_\mathrm{2\rightarrow1} = 1/\tau_2$, where $\tau_i$ are the
lifetimes of the potentials, and is submitted to state-dependent
friction, $\alpha_i \dot{x}$, and diffusive, $\xi_i(t)$, forces.  This
can be summed up in the Langevin equation of motion~\cite{Fokker2}
\begin{equation}
\ddot{x} = - \nabla_x U_i(x) - \alpha_i\dot{x} + \xi_i(t) ,
\label{eq:Langevin}
\end{equation}
where $\xi_i(t)$ is a Gaussian white noise satisfying the relations
$\langle\xi_i(t)\rangle = 0$ and $\langle\xi_i(t)\xi_i(t')\rangle =
2\alpha_i T_i \delta{(t-t')}$, using a system of units where the mass
of the Brownian particle $m=1$ and Boltzmann's constant $k_\mathrm{B}
= 1$.  The index $i \in \left\{ 1,2 \right\}$ is either a random
variable or changes at fixed times, with a rate of change set by the
transfer rates mentioned above.  When it is random, we consider the
change of state as being a source of noise, as it leads to
non-deterministic, noisy dynamics, in addition to $\xi_i(t)$ which we
term thermal noise.

The numerical integration of (\ref{eq:Langevin}) is done with the BBK
method~\cite{BBK}.  We use time steps of size 0.01, random initial
positions and velocities, and run the simulations for 1000 or 2000
particles for about 40000 time steps.

\begin{figure}
\centering
\includegraphics[width=0.6\textwidth]{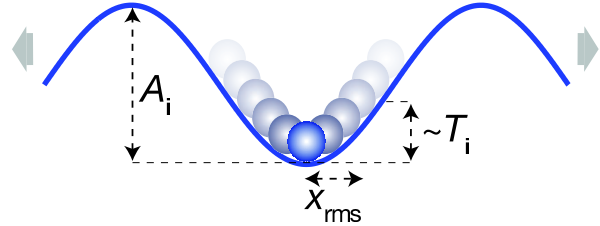}
\caption{Schematic illustration of the distribution of a Brownian
  particle in a periodic potential with friction and diffusive forces
  present. At equilibrium, the distribution will be centred around
  potential minima with a velocity spread proportional to the
  equilibrium temperature $T$, and a position spread,
  $x_\mathrm{rms}$, dependent on the temperature $T$ and potential
  depth $A$. Even though the thermal energy is less than the potential
  depth, the spread of the distribution will lead to a diffusion in
  the potential, indicated with thick horizontal arrows in the figure.}
\label{mod}
\end{figure}

The relative magnitude of the parameters are chosen to achieve a
Brownian motor and, more specifically, to mimic the experimental
system used in~\cite{OurBM1,newBM,0295-5075-81-3-33001,rtbm}. This means
that the diffusive force should be large enough to clearly influence
the dynamics, while small enough for the trapping potential to be
relevant.  As a default setting, we take $T_1=T_2=1$, $A_1=A_2=4$, and
$\varphi = 2\pi/3$ (the choice of the latter is explained in
section~\ref{sec:rates}). This gives a potential depth that is four
times larger than the thermal energy, $A_i/T_i= 4$, such that the
particles are generally localised close to the potential minima but
occasionally diffuse to neighbouring wells. A schematic illustration
of the distribution of a Brownian particle in these conditions is
shown in figure~\ref{mod},
with the width of the distribution characterised by the
root-mean-square position (or standard deviation) $x_\mathrm{rms}
\equiv \sqrt{\left\langle x^2 \right\rangle - \left\langle x
  \right\rangle^2}$ (the position is to be understood as taken modulo
$2\pi / k$, with $\left\langle \right\rangle$ denoting an ensemble
average).

\begin{figure}
\centering
\includegraphics[width=0.35\textwidth]{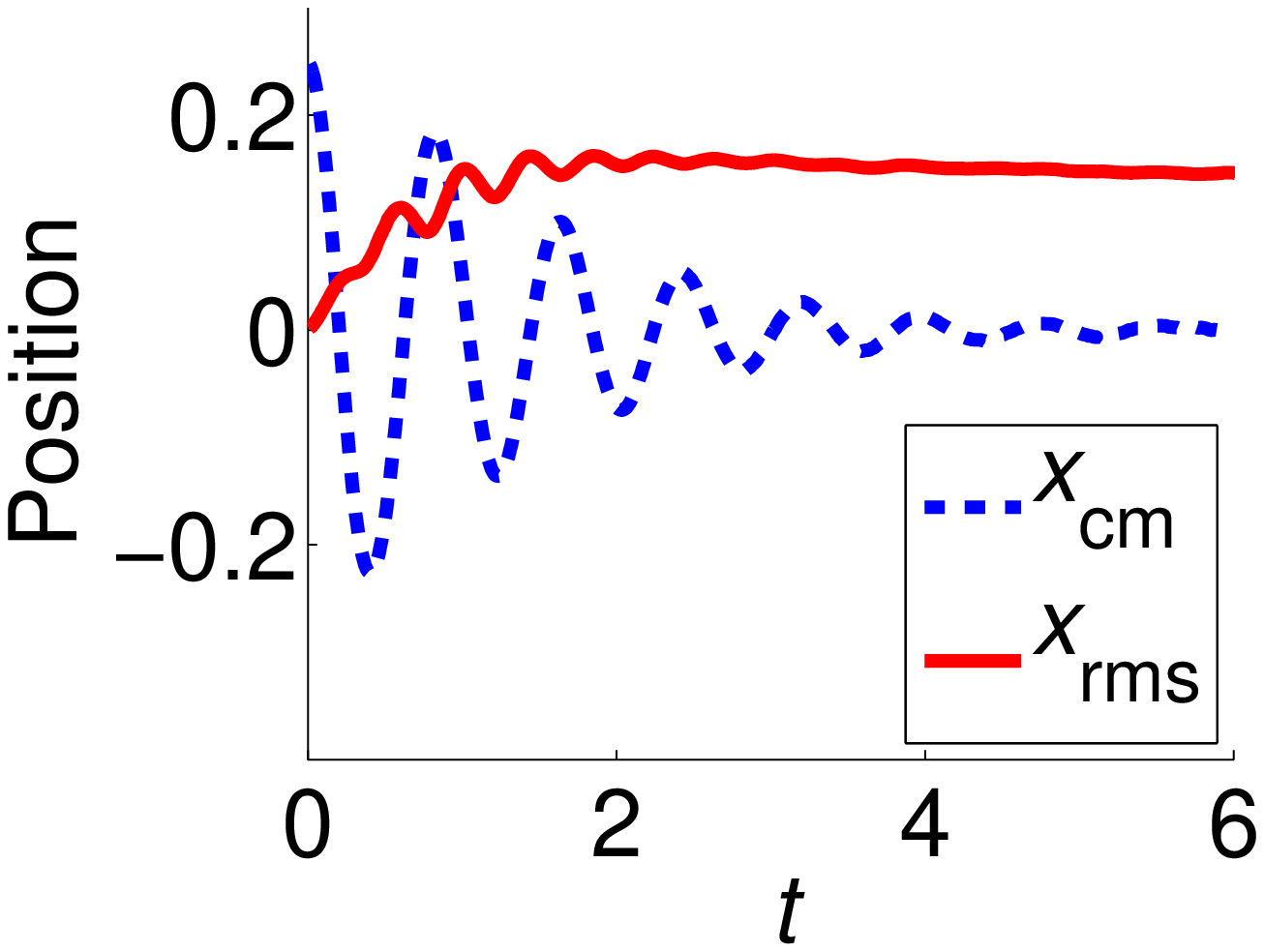}
\includegraphics[width=0.35\textwidth]{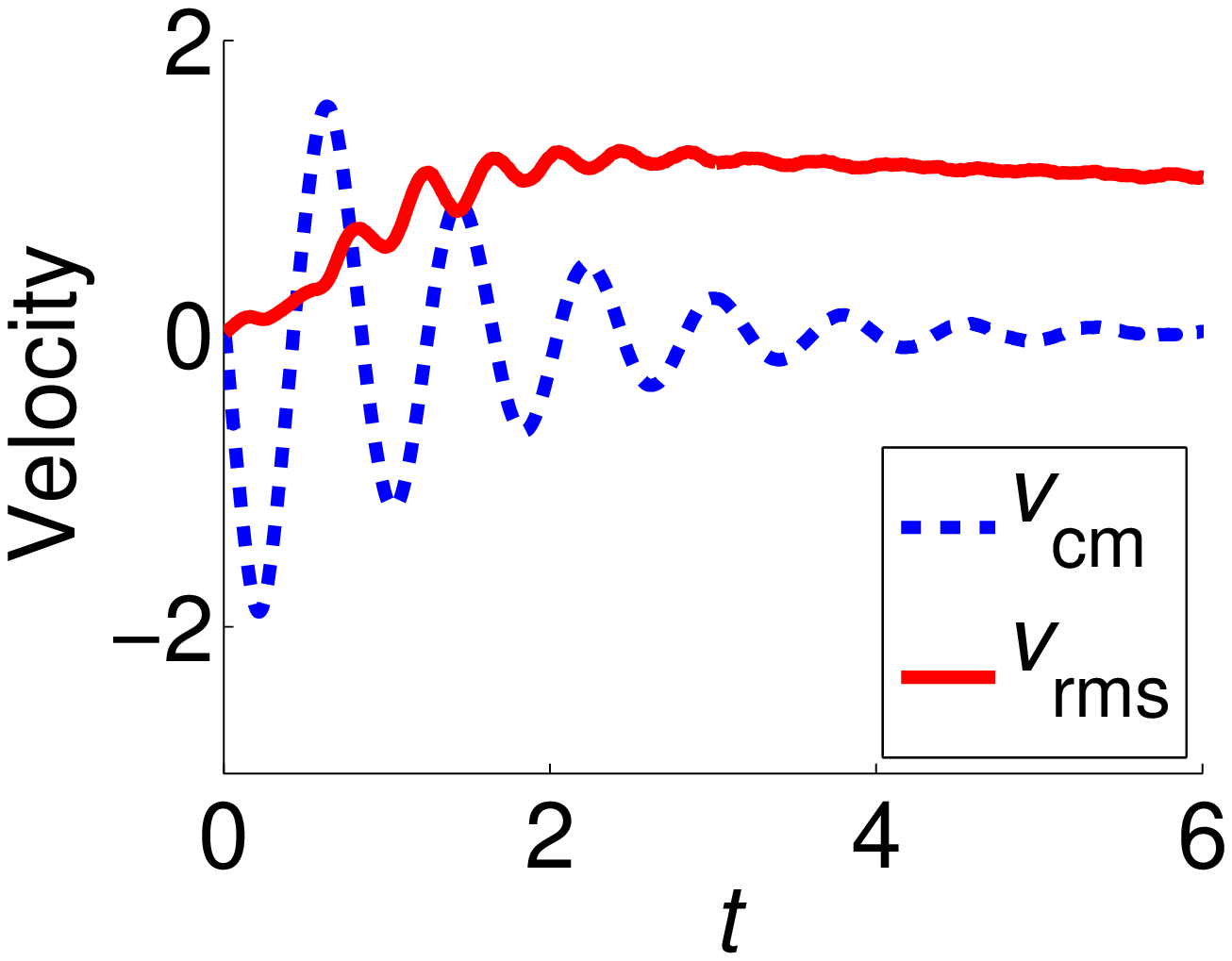}
\caption{Time evolution of the position and velocity distribution of a
  Brownian particle starting on a slope of a periodic potential with a
  finite friction and diffusive force. The initial oscillations of
  $x_\mathrm{cm}$ and $v_\mathrm{cm}$ are damped on the same time
  scale as the widths of the distributions, $x_\mathrm{rms}$ and
  $v_\mathrm{rms}$, reach their equilibrium values.}
\label{osc}
\end{figure}

The simulations are done for moderate friction, with default values of
$\alpha_{1} =\alpha_{2} =1/2$, such that the system is not overdamped,
in contrast to most other investigations of Brownian motor systems. The
equilibrium properties are similar for both underdamped and overdamped
systems, but the route to equilibrium differs significantly. This can
be illustrated by considering a moderately damped particle starting at
the slope of the potential. With time, the position and velocity
spread will increase toward the equilibrium values, see
figure~\ref{osc}. 
The average position, $x_\mathrm{cm} \equiv \left\langle x
  \pmod{2\pi/k} \right\rangle$, and velocity $v_\mathrm{cm} \equiv
\left\langle v \right\rangle$ will, on the same time scale, perform a
damped oscillation around the equilibrium value of zero ($x=0$ being
the centre of the well), see figure~\ref{osc}. In this case, the
particle could be said to gradually lose memory of earlier positions
or velocities.  For an overdamped system, no such oscillations take
place, as any memory of earlier positions or velocities are quickly
lost.

The relation between the inter-potential transfer rates and the time
to reach equilibrium crucially influences the distribution of
positions at the instant of the transfer between the potentials. To study the
effects of the oscillations before equilibrium is reached (with
friction $\alpha=1/2$), transfer rates $\Gamma_{i\rightarrow j}>1/4$
should be used, while for studies under equilibrium conditions
$\Gamma_{i\rightarrow j}\leq1/4$ be used, see figure~\ref{osc}.  The
inter-potential transfer can be done in two ways: either at fixed
times, analogous to the flashing ratchet model, or with random
transfer times, adding another source of noise to the Brownian
system. The difference in the resulting dynamics will be investigated
and discussed later in this paper.

When the transfer times are relatively short compared to the time
required to reach equilibrium in a given state (figure~\ref{osc}), the
actual kinetic temperature of the particles in state $i$ will be
greater than the equilibrium temperature $T_i$.  Therefore, when we
discuss the equilibrium properties in this case, we mean the
equilibrium that would be reached in absence of the fast transfers,
not the steady-state properties of the system.

\section{Directed transport with asymmetric transfer rates}%
\label{sec:rates}

\begin{figure}
\centering
\includegraphics[width=0.5\textwidth]{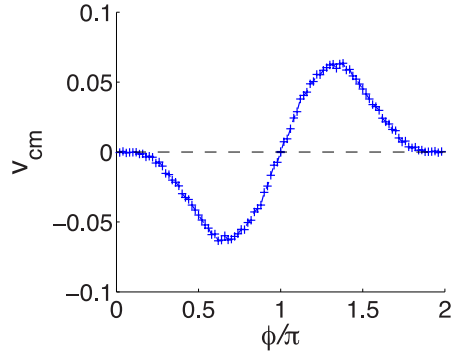}
\caption{Effect of the relative spatial phase of the two potentials on
  the average drift velocity of the particles.  The potentials are
  identical, but with unequal transfer rates between them,
  $\Gamma_\mathrm{1\rightarrow2} = 10
  \Gamma_\mathrm{2\rightarrow1}=1$.}
\label{v_phi}
\end{figure}

A necessary condition for directed transport is for the two potentials
to be out of phase (\emph{i.e.}, $\varphi \neq n \pi$, with $n$ an
integer), but this is not sufficient to completely break the symmetry
of the system.  Indeed, it was shown that no drift can be induced if
such a system is overdamped~\cite{brown:kanada99}.  Otherwise, unequal
transfer rates ($\Gamma_\mathrm{1\rightarrow2} \neq
\Gamma_\mathrm{2\rightarrow1}$) will result in directed transport, as
shown with both simulations~\cite{bm:sanchez-palencia04} and
experiments~\cite{OurBM1,newBM,0295-5075-81-3-33001,rtbm}. In both these
cases, the times at which the particle changes state are random.  We
show in figure~\ref{v_phi} the dependence of the average transport
velocity on the relative spatial phase for the model described above
with $\Gamma_\mathrm{1\rightarrow2} = 10
\Gamma_\mathrm{2\rightarrow1}=1$.
We point out that it is this case (unequal transfer rates and random
transfer times) that was assumed to most closely resemble the
experimental implementation with cold caesium
atoms~\cite{OurBM1,newBM,0295-5075-81-3-33001,rtbm}, and was thus the only
case studied theoretically in previous
work~\cite{bm:sanchez-palencia04,newBM,0295-5075-81-3-33001,BMsim}.

To better understand the dynamics of the induced drifts, we set the
relative spatial phase to the value that optimises the drift in the
previous case, $\varphi=2\pi/3$ (figure~\ref{v_phi}), and now consider
the effect of a varying asymmetry in the transfer rates. This is done
below for both fixed and random transfer times.

\subsection{Fixed transfer times}

\begin{figure}
\centering
\includegraphics[width=0.5\textwidth]{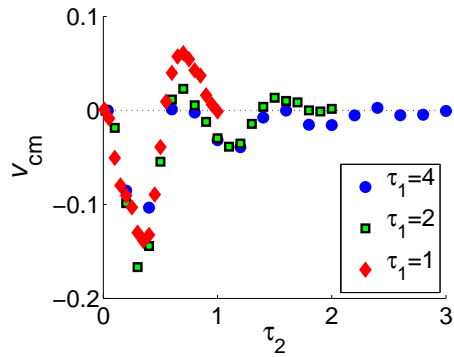}
\caption{Directed transport for fixed asymmetric transfer times,
  plotted as the centre-of-mass velocity $v_\mathrm{cm}$ \emph{vs} the
  lifetime in potential 2, $\tau_2$, for three different values of the
  lifetime in potential 1: $\tau_1 = 1$ (diamonds), $\tau_1 = 2$
  (squares), and $\tau_1 = 4$ (circles).}
\label{T0fixt}
\end{figure}

\begin{figure}
\centering
\includegraphics[width=0.4\textwidth]{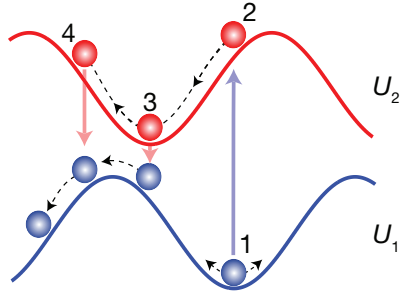}
\caption{Schematic illustration of the directed transport mechanism
  for unequal transfer rates. (1) The particle has spent a long enough
  time in potential 1 to be close to equilibrium and be well localised
  close to the bottom of a well. (2) The particle gets transferred to
  a slope in potential 2, and starts moving. (3) The particle has been
  accelerated to the bottom of a well in potential 2 and gained
  momentum. If the particle is transferred back to potential 1, this
  momentum may be enough to overcome the barrier to the next well. (4)
  If the particle moves to the other slope of the well in potential 2
  before being transferred back to potential 1, the barrier to the
  next well in potential 1 has now been passed.}
\label{bbb}
\end{figure}

By letting the transfers occur at fixed times, a single source of
randomness remains, that due to $T>0$, and the system is somewhat
simplified.  The asymmetry is here scanned by varying $\tau_2$ from
about $0.01 \tau_1$ (highly asymmetric) to $\tau_1$ (symmetric), for
three different values of $\tau_1$.  The resulting average velocity of
the induced drifts can be seen in figure~\ref{T0fixt}.
Since the two states are identical, apart from the transfer rates,
they have equal equilibrium properties. The drifts are hence a direct
consequence of the underdamped oscillations seen in
figure~\ref{osc}. These can induce drifts in two related ways, as
illustrated in figure~\ref{bbb}. 
Consider a particle located close to the bottom of a well of potential
1. As it is transferred to potential 2, it will end up on a slope, and
start moving towards the minimum. If the particle moves past the
minimum and up on the other slope before being transferred back, it
can go beyond the point of the potential barrier of potential 1.  Even
if it comes back to the same well in potential 1, the momentum gained
can be enough to overcome the barrier in potential 1.

In figure~\ref{T0fixt}, a clear oscillatory pattern in the drift
velocity's dependence on $\tau_2$ is seen. The oscillations have the
same periodicity as in figure~\ref{osc}, and their amplitude follows
the same decaying envelope. For $\tau_1=4$, four peaks in the drift
velocity can be seen, all in the same direction. For smaller $\tau_1$,
velocity peaks in the other direction can be seen, since the
pre-equilibrium oscillations are now large enough to matter in
potential 1, and the average drift hence gets a complex dependence on
the oscillations in both potentials.

These drifts are a consequence of the finite damping and this scheme
would hence not work in an overdamped
case~\cite{brown:kanada99}. Nevertheless, clear similarities with the
overdamped flashing and rocked ratchets can be seen, where large
increases in the diffusion and the drift have been reported for
certain resonances in the flashing
rates~\cite{0295-5075-44-4-416,1367-2630-11-10-103017,doi:789903}.
For an asymmetry built on unequal transfer rates with deterministic
transfer times, this works by matching transfer times and oscillation
frequencies. In the deterministic limit, as the position spread of the
particles is small, unidirectional drifts can thus be achieved.  This
situation is similar to a conventional internal combustion engine,
where the timing of the ignition with the position of the piston is
crucial, and as such the system does not correspond \emph{per se} to a
Brownian motor.

\subsection{Random transfer times}

\begin{figure}
\includegraphics[width=0.45\textwidth]{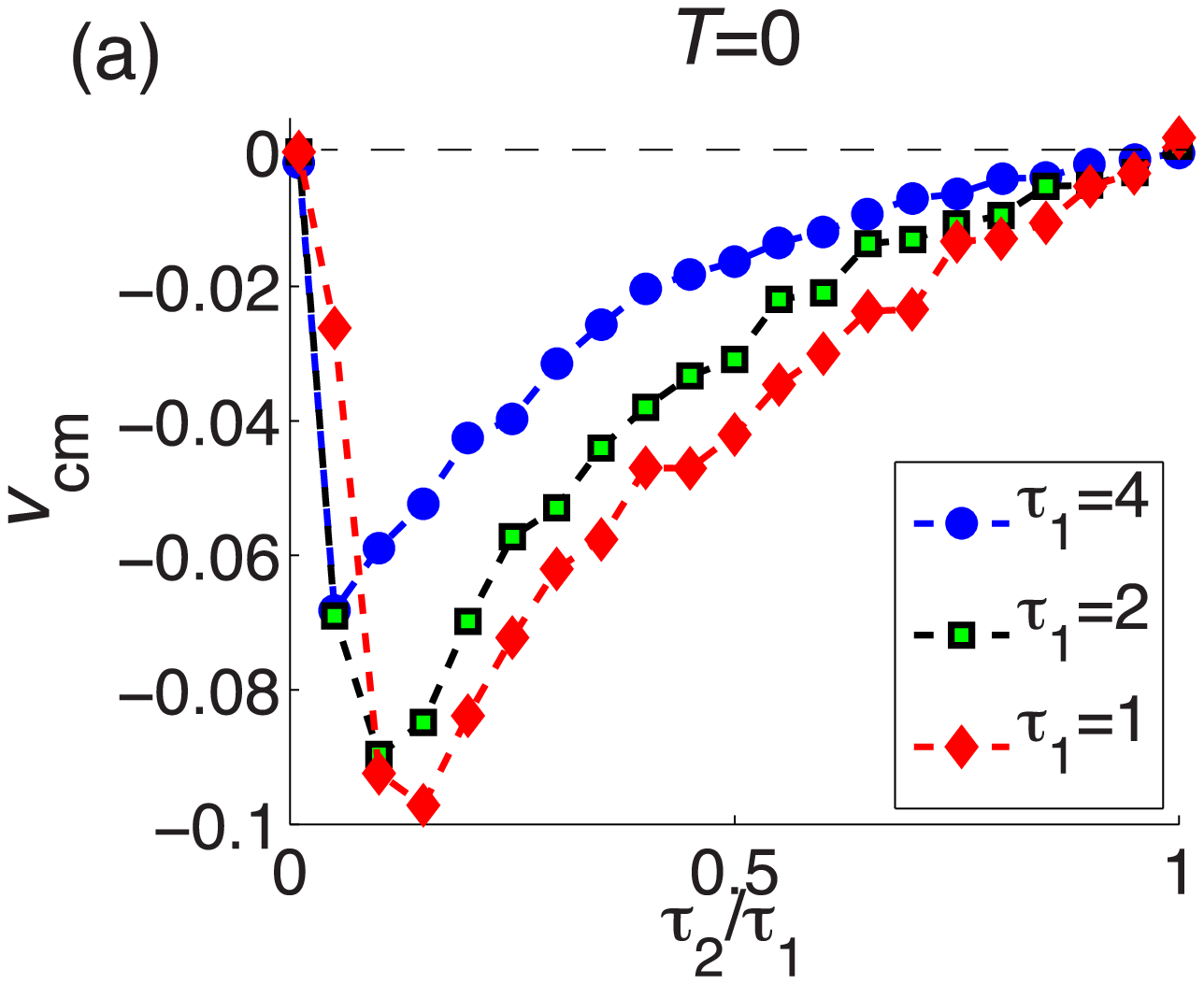}
\includegraphics[width=0.45\textwidth]{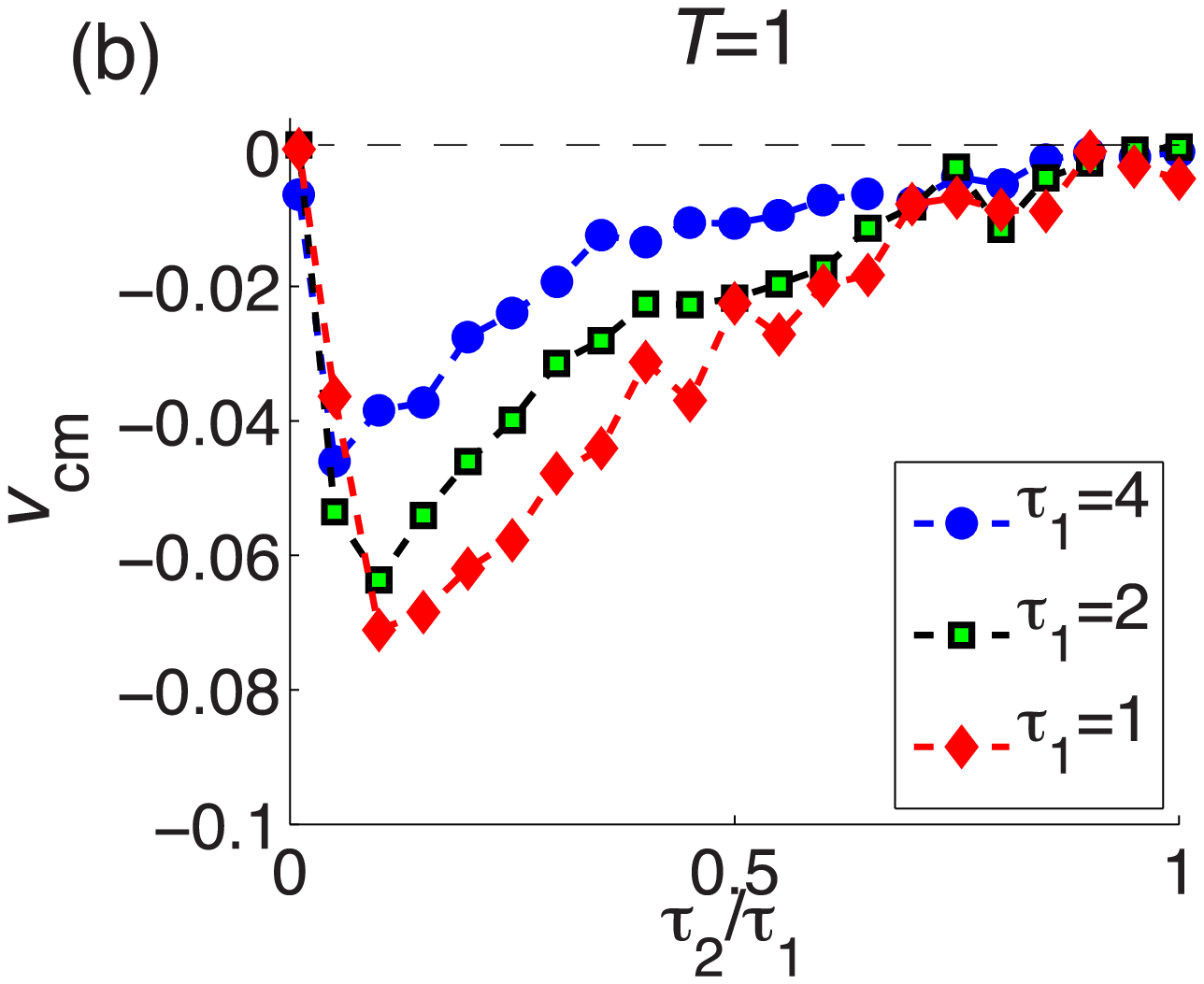}
\caption{Directed transport for random asymmetric transfer times,
  plotted as the centre-of-mass velocity $v_{cm}$ \emph{vs} the
  lifetime in potential 2, $\tau_2$, for three different values of the
  lifetime in potential 1: $\tau_1 = 1$ (diamonds), $\tau_1 = 2$
  (squares), and $\tau_1 = 4$ (circles). (a) $T=0$, showing drifts for
  all $\tau_1 \neq \tau_2$. (b) $T=1$, resulting in the same structure
  as for $T=0$, but with slightly smaller drifts.  The lines serve as
  a guide to the eye.}
\label{Trand}
\end{figure}

Here the inter-potential transfers occur at random times, adding a
source of randomness to the dynamics.  With these extra fluctuations
present, the average drift velocity is monitored for a varying average
lifetime in potential 2, $\tau_2$, for three different average
lifetimes in potential 1, $\tau_1$. The results are seen in
figure~\ref{Trand}. To distinguish the effects of the two noise
sources, the simulations are done for both $T=0$ and $T=1$.

All resonance behaviour seen in figure~\ref{T0fixt} has disappeared,
and all drifts are induced in the same direction, even for short
transfer times. That is, the drifts are induced due to asymmetries in
position or velocity space rather than by transfers at deterministic
positions. In other words, since the transfer times are now a
distribution rather than specific times, there is an uneven
competition between processes that lead to diffusion in both
directions.  The similarity to conventional engines is lost and the
behaviour is that of a ``true'' Brownian motor. Note that similar
results are achieved both with and without the thermal noise, and that
the equilibrium noise slows down the drifts rather than enhance them.
However, random transfer times eliminate the need for a precise timing
of the transfers.

Since it is the noise from the random transfers that is rectified,
rather than the thermal noise, drifts can be induced for particles
that are initially at rest. This is in clear contrast to the flashing
ratchet, where the drift comes from an enhancement of the thermal
noise. For a flashing ratchet, zero velocity corresponds to an
absorbing state for $T=0$, even if random flashing would be used.

Since the equilibrium states are equal, the drifts are a consequence
of the finite damping, as was also the case for fixed transfer times,
and no drifts would be induced if the motion became
overdamped~\cite{brown:kanada99}.  We will now explore the possibility
of inducing drifts with equal transfer rates but unequal equilibrium
properties in the two states.

\section{Breaking the symmetry of time-symmetric potentials}%

We now consider induced drifts with equal transfer rates,
$\Gamma_\mathrm{1\rightarrow2} = \Gamma_\mathrm{2\rightarrow1}$, by
introducing inequalities in the equilibrium states.  The equilibrium
states have zero average velocity and an average position centred in
the wells, see figures~\ref{mod} and \ref{osc}.  Therefore, no
resonances in the timing of the transfer can be found after the
equilibrium has been reached. The equilibrium states are characterised
by the widths of the velocity and position distributions in each
well. In our model, these depend on the temperatures and the potential
depths of the two states. We now investigate the effects of
asymmetries in these parameters.

\subsection{States with different temperatures}
The temperature determines the width of the equilibrium distribution
of both the velocity and position within a well. In potentials of
equal depth, different temperatures thus result in differences between
the distributions in the two potentials. This, together with a
relative spatial phase, breaks the symmetry and can hence induce
drifts.

\begin{figure}
\centering
\includegraphics[width=0.45\textwidth]{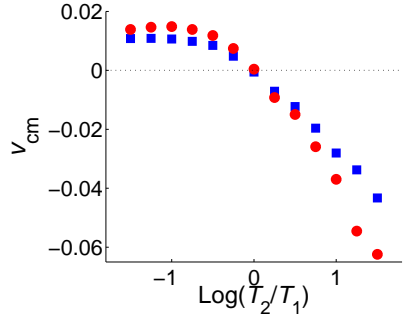}
\caption{Drift velocity resulting from unequal temperatures in the two
  potentials, for random (squares) and fixed (circles) transfer times
  ($\tau_1 = \tau_2 = 10$). $T_1=1$ and $T_2$ is varied.}
\label{Tvar}
\end{figure}

To demonstrate this, $T_2$ is varied for fixed $T_1=1$, in otherwise
identical potentials. Since it is the equilibrium properties that are
interesting, transfer times long enough to reach equilibrium are used,
$\tau_1 = \tau_2 = 10$. The results for both deterministic and random
transfer times are seen in figure~\ref{Tvar}.
Slightly higher drift velocities for deterministic transfer times are
obtained. This is expected since all transfers occur when the two
unequal equilibrium states have been reached for the case of
deterministic transfers, while for random times a few transfers occur
before equilibrium has been reached, which is on average unfavourable.

\begin{figure}
\centering
\includegraphics[width=0.45\textwidth]{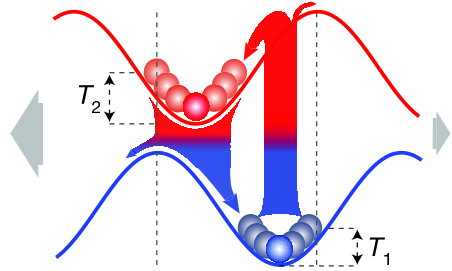}
\caption{Schematic representation of different temperatures in the two
  potentials. The spread of the position distributions together with
  the relative spatial phase will lead to an increased diffusion of
  the particles due to the transfers. The unequal widths of the
  distributions bias this diffusion in one direction.}
\label{Tvar_mod}
\end{figure}

\begin{figure}
\centering
\includegraphics[width=0.45\textwidth]{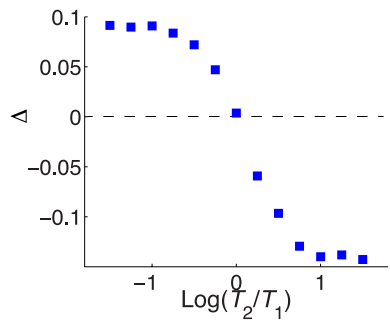}
\caption{Asymmetry in the probability of a particle ending up in a
  neighbouring well, $\Delta$ (\ref{eq:delta}), resulting from unequal
  temperatures in the two potentials ($T_1=1$ and $T_2$ is varied).}
\label{dg}
\end{figure}

The symmetry breaking can here be illustrated by considering the
equilibrium distributions in the two potentials, see
figure~\ref{Tvar_mod}. Just before a transfer, the position
distribution will be centred around a minimum in potential 1 with a
certain spread. If the relative spatial phase is non-zero and less
than $\pi$, the position of a minimum in potential 1 is located in
between a minimum and a maximum of potential 2, see
figure~\ref{Tvar_mod}.  If the spread is sufficiently small, the
particle will, with a probability close to unity, end up after the
transfer on the slope pointing toward the closest minimum in potential
2. However, if the spread is large enough, there is a finite
probability to end up on the slope of the neighbouring minimum,
leading to a diffusion of the particles. If the distributions in the
two potentials have unequal spreads, the increased diffusion will be
biased in one direction. To further investigate the mechanism of this
biased diffusion, we define the difference between the probability of
ending up in a neighbouring minimum,
$P^\mathrm{neighbour}_{i\rightarrow j}$, for the two transitions as
\begin{equation}
  \Delta \equiv   P^\mathrm{neighbour}_{1\rightarrow 2} -
  P^\mathrm{neighbour}_{2\rightarrow 1} .
\label{eq:delta} 
\end{equation}
The value of $\Delta$ can be extracted from the simulations by keeping
track of the positions at which the transfers occur.
If $\Delta$ is plotted for the same asymmetries as in
figure~\ref{Tvar}, the same basic dependency is recreated, see
figure~\ref{dg}. A small deviation is visible for large
temperatures. This could either be due the states having different
velocity distributions or, since the finite damping allows for
multi-well flights, a difference in the average flight length in each
direction.

In contrast to the case of unequal transfer rates, it is here the
thermal noise, which creates the spread of the equilibrium
distributions, that is enhanced and rectified, just as for a flashing
ratchet. This type of symmetry breaking would hence also work for an
overdamped system. Drifts from two heat baths with different
temperatures are analogous to the Feynman-Smoluchowski thought
experiment~\cite{rat_praw}.

\subsection{States with different potential depths}
The symmetry can be broken by unequal potential depths as well.  While
this is formally equivalent to having different temperatures
corresponding to each state, as both cases correspond to unequal
energy scales for the two states, treating it independently allows us
to make clearer the role of the thermal noise.

\begin{figure}[htbp]
\centering
\includegraphics[width=0.45\textwidth]{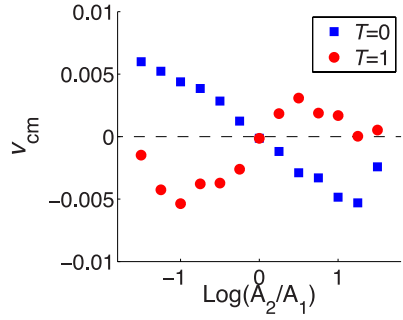}
\caption{Drift velocity resulting from unequal potential depths
  ($A_1=4$ and $A_2$ is varied), for $T=0$ (squares) and $T=1$
  (circles) with random transfer times $\Gamma_{1 \rightarrow 2} =
  \Gamma_{2 \rightarrow 1}= 1/10$. The equilibrium noise ($T$)
  inverses the average drift velocity's dependence on $A_2 / A_1$.}
\label{varA}
\end{figure}

\begin{figure}
\centering
\includegraphics[width=0.35\textwidth]{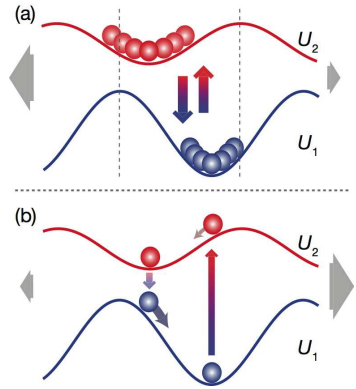}
\caption{Schematic representation of potentials with different
  depths. (a) The spread of the position in the equilibrium
  distributions together with the relative spatial phase will lead to
  an increased diffusion of the particles. Due to the unequal widths
  of the distributions, this diffusion will be biased in one direction
  for long transfer times. (b) The difference in potential depths
  gives different slopes and different velocities for short transfer
  times. The two effects and (a) and (b) have opposite directions.}
\label{T_mod}
\end{figure}

As with temperature, the potential depth affects the equilibrium
distribution in position: a larger depth gives a narrower spread in
position.  In figure~\ref{varA}, the drift velocity is shown for a
varying depth of potential 2. This is done for random transfer times,
and for both for $T=0$ and $T=1$.  To allow the system to reach
equilibrium between transfers, the transfer times are again set to
$\tau_1 = \tau_2 = 10$.
Note that the dependence of the drift on $A_2 / A_1$ is inverted for
the two temperatures, which can be explained as follows.  For finite
but equal temperatures, the position spread will be greater for the
shallower potential and, for the current set of parameters, this leads
to a drift in the negative direction when $A_2 < A_1$ (see
figure~\ref{T_mod}(a)) and \emph{vice versa}.  When $T=0$, if the
transfer times are short enough, the dynamics are dominated by the
velocity acquired when changing potential, which is greater when going
to the deeper potential, as illustrated in figure~\ref{T_mod}(b),
resulting a drift in the positive direction for $A_2 < A_1$.  For
longer transfer times, where the system would have to reach
equilibrium between transfers, no drift would be observed for $T=0$.
This phenomenon of purely noise-induced reversal of the drift has also
been reported for rocked ratchets~\cite{0295-5075-44-4-416}.

\begin{figure}
\centering
\includegraphics[width=0.45\textwidth]{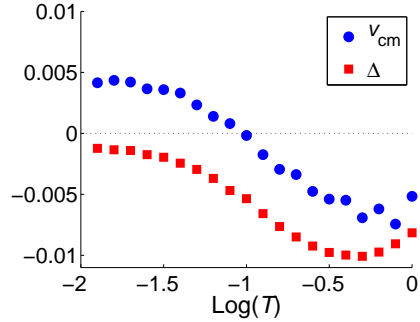}
\caption{Drift velocity $v_\mathrm{cm}$ and probability asymmetry
  $\Delta$~(\ref{eq:delta}) for different temperatures ($T_1 = T_2 =
  T$), for $A_2/A_1=1/10$.}
\label{Trev}
\end{figure}

To further investigate the noise-induced reversal of the drift, the
temperatures are scanned from 0.01 to unity, for an asymmetry
$A_2/A_1=1/10$ (resulting large drifts in both directions, see
figure~\ref{varA}). The drift achieved is shown in figure~\ref{Trev},
together with $\Delta$~(\ref{eq:delta}), where a
correlation between the two is evident.

With fixed transfer times, the dominant effect, and thus the direction
of the drift, depends on the value of the transfer rate.  This is only
possible for the underdamped case, where resonances due to transient
effects can be found, as for an overdamped system only the equilibrium
properties are of importance.

\subsection{States with different friction}

\begin{figure}
\centering
\includegraphics[width=0.45\textwidth]{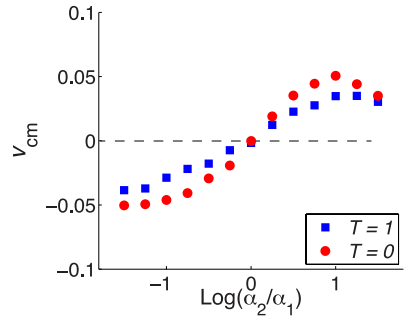}
\caption{Drift velocity for different friction coefficients in the two
  potentials for $T=0$ (circles) and $T=1$ (squares). $\alpha_2$ is
  varied for constant $\alpha_1 = 5$.}
\label{fric_var}
\end{figure}

In the Langevin formulation of the problem~(\ref{eq:Langevin}),
friction does not change the temperature of the equilibrium states
(indeed, friction can be removed from the equations by the rescaling
$t \rightarrow \alpha t$).  What the friction does change is the decay
time of the pre-equilibrium oscillations (figure~\ref{osc}). Unequal
frictions can hence induce drifts, and in figure~\ref{fric_var} drifts
in both directions are evident for a varying
$\alpha_2/\alpha_1$. 
Since the drifts do not depend here on unequal
equilibrium states, this scheme works for both $T=0$ and $T=1$.

\begin{figure}[htbp]
\centering
\includegraphics[width=0.45\textwidth]{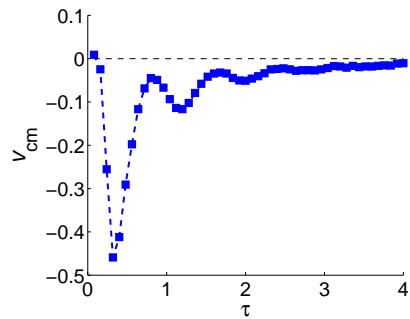}
\caption{Drift velocity $v_{cm}$ \emph{vs} the deterministic transfer
  time $\tau$ of both potentials for a friction inequality of
  $\alpha_1 = 5$ and $\alpha_2 = 1/2$. The line serves as a guide to
  the eye.}
\label{fric_var2}
\end{figure}

This asymmetry is closely related to the case of unequal transfer
rates. This is clearly seen by using fixed equal transfer times,
$\tau_1=\tau_2=\tau$, and scanning these for a friction inequality
$\alpha_1=5$, $\alpha_2=1/2$, which gives large drifts, see
figure~\ref{fric_var2}.

\section{Conclusion}

We have shown that two-state Brownian motors, realised with Brownian
particles in symmetric potentials, are very rich systems, where the
necessary spatio-temporal asymmetry required for a directed motion can
be induced by a differentiation of either the two potential depths,
the two temperatures, the two frictions, or the two lifetimes of the
system.  For the case of different transfer rates or different
frictions, the random fluctuations associated with the random
inter-potential transfer are shown to be the energy source,
\emph{i.e.}, the noise that is rectified. For the case of different
potential depths or different equilibrium temperatures, it is the
thermal noise in each potential that was shown to be the relevant
energy source.  For the case of an asymmetry in the potential depth, a
pure noise-induced reversal of the drift is also demonstrated.

The richness of this Brownian motor makes it flexible and clear
similarities to several different types of systems can be seen,
depending on the source of the asymmetry.  It serves as a good
exemplification of the statement of Pierre Curie that ``when certain
causes produce certain effects, the elements of symmetry of the causes
must be found in the effects produced''~\cite{Curie}, as in our system
\emph{any} asymmetry will lead to a directed motion.  The results
demonstrate also that the source of the symmetry breaking of a
Brownian motor can be diverse, and indeed even many asymmetries can be
present simultaneously, which would typically be the case for atoms in
a double optical
lattice~\cite{OurBM1,newBM,0295-5075-81-3-33001,rtbm}.  The increased
understanding of the model as such also increases the understanding,
and possibilities to optimise, the above-mentioned existing
experimental realisation of the system.

\ack
C.M.D. gratefully acknowledges funding from the Swedish Research
Council and Ume{\aa} University.

\section*{References}

\end{document}